\documentclass[universe,article,accept,moreauthors,pdftex]{mdpi}

\firstpage{1} 
\pubvolume{8}
\issuenum{1}
\articlenumber{302}
\pubyear{2022}
\copyrightyear{2020}
\externaleditor{Academic Editors: Antonino Del Popolo, Yi-Fu Cai and Jean-Michel Alimi} % For journal Automation, please change Academic Editor to "Communicated by"
\datereceived{19 march 2022} 
\dateaccepted{25 may 2022}
\datepublished{26 may 2022}
\hreflink{https://doi.org/\\10.3390/universe8060302}

\usepackage[english]{babel}
\usepackage{dcolumn}% Align table columns on decimal point
\usepackage{bm}% bold math
\usepackage{verbatim}   % useful for program listings
\usepackage{color}      % use if color is used in text
\usepackage{mathrsfs}
\usepackage{braket}
\usepackage{physics}
\usepackage{amssymb}
\usepackage{subfig}
\usepackage{float}
\usepackage{cancel}
\usepackage{bigints}
\usepackage{trackchanges}
\Title{Polymer dynamics of isotropic universe in Ashtekar and in volume variables}
\TitleCitation{Polymer dynamics of isotropic universe in Ashtekar and in volume variables}
\Author{Eleonora Giovannetti $^{1,\dagger}$\orcidA{}*, Gabriele Barca $^{1,\dagger}$\orcidB{}, Federico Mandini $^{1,\dagger}$ and Giovanni Montani $^{2,1,\dagger}$\orcidC{}}
\AuthorNames{Eleonora Giovannetti, Gabriele Barca, Federico Mandini and Giovanni Montani}
\AuthorCitation{Giovannetti, E.; Barca, G.; Mandini, F; Montani, G.}
\address{$^{1}$ \quad Department of Physics, "La Sapienza" University of Rome, P.le Aldo Moro 5 (00185) Roma, Italy\\
	$^{2}$ \quad ENEA, Fusion and Nuclear Safety Department, C.R. Frascati, Via E. Fermi, 45 (00044) Frascati (RM), Italy}
\corres{Correspondence: eleonora.giovannetti@uniroma1.it}
\firstnote{These authors contributed equally to this work.}

\abstract{We analyze the semiclassical and quantum polymer dynamics of the isotropic Universe in terms of both the standard Ashtekar-Barbero-Immirzi connection and its conjugate momentum and also of the new generalized coordinate conjugate to the Universe volume. We study the properties of the resulting bouncing cosmology that emerges in both the representations and we show that the Big Bounce is an intrinsic cut-off on the cosmological dynamics only when the volume variable is implemented, while in terms of the standard connection the Universe Bounce energy density is fixed by the initial conditions on the prepared wavepacket. As a phenomenological implication, we introduce particle creation as a dissipative term and study the production of entropy in the two formulations. Then, we compare the obtained dynamics with what emerges in Loop Quantum Cosmology, where the same difference in the nature of the Big Bounce is associated to fixing a minimum area eigenvalue in a comoving or in a physical representation. We conclude that the privileged character of the Ashtekar-Barbero-Immirzi connection suggests that the natural scenario in the polymer framework is a Big Bounce that is not a Universal cut-off. However, by a parallelism between the polymer and Loop Quantum Cosmology properties of the basic operators, we also develop some considerations in favour of the viability of the $\bar{\mu}$ scheme of Loop Quantum Cosmology on a semiclassical level.}


\begin{document}
	
\section{Introduction}
One of the most intriguing implications of Loop Quantum Gravity (LQG) \cite{ROVELLI199080,Rovelli_1995DiscretenessAreaVolume,QuantumGeomI,QuantumGeomII} (for recent reviews, see \cite{LQG30years,LQGReview_2021}) is the emergence of a bouncing cosmology in the reduced model obtained when the symmetries of the cosmological principle are implemented. Such a formulation of the full theory within a minisuperspace scenario is commonly dubbed Loop Quantum Cosmology (LQC) \cite{BojowaldOriginalLQC,Ashtekar0_2006,Ashtekar_2006,Ashtekar2_2006,Ashtekar_2011review} (for the anisotropic extension of this framework see \cite{Wilson-Ewing_BianchiI,Wilson-Ewing_BianchiII,Wilson-Ewing_BianchiIX,Wilson-Ewing_BianchiIcomplete,Wilson_Ewing_2016}) and it offers a non-singular framework to implement the cosmological history of the Universe (actually, after the Planckian time the Universe thermal history remains faithful to the original formulation \cite{PhysRev.74.505.2,Weinberg:1972kfs,Kolb:1990vq,Montani:2011zz}).
	
However, the minisuperspace implementation of LQG has the non-trivial limitation that the basic $SU(2)$ symmetry is essentially lost and the discretization of the area operator spectrum is somewhat introduced ad hoc, in contrast with LQG where it takes place naturally on a kinematical level \cite{Ashtekar_2006,Ashtekar2_2006}. The difficulties of LQC in reproducing the fundamental character of the general quantum theory have been discussed in \cite{CIANFRANI_2012}, and alternative Loop quantization procedures have been proposed in \cite{Bojowald_2009Consistent,Alesci1,Alesci2,Alesci3,Alesci4}; for a more thorough criticism of the whole cosmological setting of LQG see \cite{Bojowald_2020}.
	
In this paper we face a specific question on the nature of the Bounce and we do it in the framework of Polymer Quantum Mechanics (PQM) \cite{Corichi_2007,CorichiHilbertSpace,BarberoPoly} implemented on the isotropic Universe. We analyze the resulting cosmological dynamics to make a comparison with the properties of the bouncing cosmology that emerges in LQC. In particular, we study the polymer quantum dynamics of Friedmann-Lemaitre-Robertson-Walker (FLRW) model both when the basic variables are the Ashtekar connection and its conjugate momentum (the flux operator) and also when the addressed phase-space variables are the new generalized coordinate and its conjugate momentum (actually the volume-like variable). We see that when the natural gauge connection is considered the density cut-off depends on the energy-like eigenvalue (i.e. on the initial conditions for a given wavepacket) as in the $\mu_0$ scheme of LQC \cite{Ashtekar_2006,BojowaldOriginalLQC}, while in the new set of variables the critical energy density is fixed by fundamental constants (i.e. the Immirzi parameter and the polymer one) as in the improved $\bar{\mu}$ scheme presented in \cite{Ashtekar2_2006,Ashtekar_2008}. Moreover, in order to make some comparison between the polymer semiclassical dynamics in the two sets of variables also on a phenomenological level, we introduce in the model an additional matter component that satisfies a continuity equation with a dissipative term, namely particle creation; this way we can analyze on a semiclassical level the different behaviour of entropy and its dependence on initial conditions.
	
The focus of the present analysis is not the existence of a bouncing cosmology in LQG and hence in LQC (which is a consequence of the inclusion of the holonomy corrections in the Hamiltonian constraint) and in Polymer Quantum Cosmology (PQC), but what is the physical interpretation of the two schemes and the link between the improved analysis performed in \cite{Ashtekar2_2006} and the original picture \cite{Ashtekar_2006}. Furthermore, the analysis performed via the improved Hamiltonian in \cite{Ashtekar2_2006} seems to be affected by an ambiguous change of variables, which is required in order to restore a standard translational operator with constant step. In fact, the Universe volume (i.e. the cubed cosmic scale factor) has its own conjugate variable corresponding to a redefined generalized coordinate which does not implement LQG features into the symmetries of the minisuperspace in the same way as with the original $SU(2)$ Ashtekar connection.
	
In this respect, from the point of view of PQM we show how, on a semiclassical level, restoring the natural Ashtekar gauge connection after the lattice has been implemented on the volume variable is formally equivalent to considering the basic lattice parameter as a function of the momentum variable; accounting for this redefinition of the polymer parameter, the Universe volume obeys the same dynamical equations in the two sets of variables. Thus, the contribution of this paper is twofold: on one hand, the polymer quantization of the isotropic Universe leads to say that, if we state that the privileged variable is the connection induced by the full theory, we arrive to a bouncing dynamics whose maximum density is not fixed a priori by fundamental constants; on the other hand, we also provide a brief argument in favor of the viability of the $\bar{\mu}$ scheme of LQC in the volume variable, by showing in the polymer framework its semiclassical equivalence to the analysis in the natural Ashtekar-Barbero-Immirzi connection.
	
The paper is structured as follows. In section \ref{POLY} we introduce the polymer representation of quantum mechanics. In section \ref{semiclassical} we apply the polymer framework to the classical dynamics of the FLRW Universe and compare the behaviour of particle creation and entropy production in the two sets of variables; in section \ref{quantum} we implement the full quantum theory and analyze the wavepacket dynamics by calculating the expectation value of the operators that best describe the evolution of this model, that are the energy density and the volume. In section \ref{discussion} we discuss and compare the results of the previous sections and argue that the two different sets of variables provide inequivalent theories; we further suggest a possibility to recover the equivalence at a semiclassical level. Then we discuss the implications of our semiclassical analysis of the equations of motion, in favour of the viability of the $\bar{\mu}$ scheme in LQC. In section \ref{concl} we conclude the paper with a brief summary and we stress some remarks. Note that for our analysis we will use $\hslash=c=8\pi G=1$.
	
\section{Polymer Quantum Mechanics}
\label{POLY}
PQM is an alternative representation that is non-unitarily connected to the standard Schr\"odinger representation. It implements a fundamental scale in the Hilbert space through the introduction of a lattice structure. For a more detailed introduction to the polymer representation see \cite{Corichi_2007}.
	
We consider a Hilbert space $\mathcal{H}'$ with the orthonormal basis $\ket {\beta_i}$ where $\beta_i \in \mathbb{R}$, $i=1,...,N$ and such that $\braket{\beta_i}{\beta_j}=\delta_{ij}$. The polymer Hilbert space $\mathcal{H}_\text{poly}$ is the completion of $\mathcal{H}'$. We can define two fundamental operators, the label operator $\hat\epsilon\ket{\beta}=\beta\ket{\beta}$ and the shift operator $\hat{s}(\zeta)\ket{\beta} =\ket{\beta+\zeta}$; the latter is actually a family of parameter-dependent unitary operators that however are discontinuous and cannot be generated by the exponentiation of a self-adjoint operator.
	
In a Hamiltonian system with canonical variables $Q$ and $P$ in the momentum polarization, states have wavefunctions $\psi_\beta(P)=\braket{P}{\beta}=e^{i\beta P}$; the two fundamental operators can be identified with the differential coordinate operator $\hat Q$ and with the multiplicative operator $\hat{T}(\zeta)=e^{i\zeta P}$:
	%\begin{subequations}
	\begin{equation}
		\hat Q \psi_\beta (P) = -i\pdv{P} \psi_\beta (P) = \beta \psi_\beta(P),\qquad\hat T(\zeta) \psi_\beta(P) = e^{i\zeta P} e^{i\beta P} = \psi_{\beta+\zeta}(P).
	\end{equation}
	%\end{subequations}
The Hilbert space becomes $\mathcal{H}_\text{poly}=L^2(\mathbb{R}_B,d\mu_H)$, the same of kinematical LQC \cite{Ashtekar_2006,BojowaldOriginalLQC}.
	
Since $\hat T (\zeta)$ is now the shift operator in $\mathcal{H}_\text{poly}$, the momentum $P$ cannot exist as the generator of translations and cannot be promoted to a well-defined operator; it must be regulated through the introduction of a lattice, i.e. a regular graph $\gamma_{\beta_0}=\{Q \in \mathbb{R} : Q=\beta_n=n\beta_0 \text{ with } n \in \mathbb{Z}\}$, where $\beta_0$ is the constant lattice spacing. The Hilbert space is then restricted to the subspace $\mathcal{H}_{\gamma_{\beta_0}} \subset \mathcal{H}_\text{poly}$ that contains all those states $\ket \psi$ such that $\ket \psi = \sum_n b_n \ket{\beta_n}$, with $\sum_n |b_n|^2 < \infty$. Now the translational operator must be restricted to act only by discrete steps in order to remain on $\gamma_{\beta_0}$ by setting $\zeta=\beta_0$: $\hat{T}(\beta_0) \ket {\beta_n} = \ket {\beta_{n+1}}$.
	
When the condition $P \ll \frac{1}{\beta_0}$ is satisfied, we can write:
	\begin{equation}
		\label{ppoly}
		P \approx \frac{1}{\beta_0} \sin{(\beta_0 P)} = \frac{1}{2i\beta_0}\bigl(e^{i\beta_0 P}-e^{-i\beta_0 P} \bigr)
	\end{equation}
and in return we can approximate the action of the momentum operator by that of $\hat T(\beta_0)$:
	\begin{equation}
		\hat P_{\beta_0} \ket{\beta_n} = \frac{\hat T(\beta_0) - \hat T(-\beta_0)}{2i\beta_0}\,\ket{\beta_n}= \frac{\ket{\beta_{n+1}} - \ket{\beta_{n-1}}}{2i\beta_0}.
	\end{equation}
For the squared momentum operator there are infinitely many approximations; we will use
	\begin{subequations}
		\begin{equation}
			\hat P^2_{\beta_0} \ket {\beta_n} = \frac{1}{4\beta_0^2}\bigl(2-\hat T(2\beta_0)-\hat T(-2\beta_0) \bigr) \ket{\beta_n},
			\label{pmu02shift}
		\end{equation}
		\begin{equation}
			P^2 \approx \frac{1}{\beta^2_0} \sin^2{(\beta_0 P)}.
			\label{p2sin2}
		\end{equation}
	\end{subequations}
Now we can implement a Hamiltonian operator on the graph as $\hat{\mathcal{C}}_{\gamma_{\beta_0}} = \frac{1}{2m} \hat P^2_{\beta_0} + \hat U(\hat Q)$, where $\hat U(\hat Q)$ is a potential.
	
When performing the quantization of a system using the momentum polarization of the polymer representation, the regulated momentum operator \eqref{pmu02shift} must be used together with the differential coordinate operator. Alternatively, it is possible to perform a semiclassical analysis by using the formal substitution \eqref{p2sin2} in the classical Hamiltonian, thus including quantum modifications in the classical dynamics.
	
\section{Polymer semiclassical dynamics of the FLRW Universe}
\label{semiclassical}
We will now apply the polymer representation to the FLRW Universe filled with matter in the form of a scalar field $\phi$. We will use both the Ashtekar variables $(p,c)$ and the volume variables $(\nu,\tilde{c})$; the latter have been shown to be the suitable variables in order to obtain a universal cut-off in polymer cosmology \cite{Montani_2019}.
	
\subsection{Dynamics in the two representations}
The polymer paradigm is implemented on the classical Hamiltonian by considering the "position" variables $p\propto a^2$ and $\nu\propto a^3$ as discrete and therefore using the substitution \eqref{p2sin2} on the conjugate momenta $c\propto\dot{a}$ and $\tilde{c}\propto\dot{a}/a$. Thus the two polymer modified Hamiltonian constraints are
	\begin{equation}
		\label{Cpoly}
		\mathcal{C}_{\text{poly}}(p,c;\phi,P_\phi)=-\frac{3}{\gamma^2\beta_0^2}\sqrt{p\,}\,\mbox{sin}^2(\beta_0c)+\rho_\phi\,p^{\frac{3}{2}}=0,\qquad\rho_\phi=\frac{P_\phi^2}{2p^3};
	\end{equation}
	\begin{equation}
		\label{CpolyV}
		\mathcal{C}_{\text{poly}}(\nu,\tilde{c};\phi,P_\phi)=-\frac{27}{4\gamma^2\beta_0^2}\,\nu\,\mbox{sin}^2(\beta_0\tilde{c})+\rho_\phi\,\nu=0,\qquad\rho_\phi=\frac{P_\phi^2}{2\nu^2};
	\end{equation}
where $\gamma$ is the Immirzi parameter \cite{Immirzi_1997,Immirzi2}, while $P_\phi$ is the momentum conjugate to the scalar field $\phi$ and is a constant of motion.
	
Thanks to the equations of motion and the Hamiltonian constraint, we find analytic expressions for the modified Friedmann equations:
	\begin{equation}
		H^2_p=\Big(\frac{\Dot{p}}{2p}\Big)^2=\frac{\rho_\phi}{3}\Big(1-\frac{\rho_\phi}{\bar{\rho}_p}\Big),\quad\bar{\rho}_p=\frac{3}{\gamma^2\beta_0^2p};
		\label{Friedmannpc}
	\end{equation}
	\begin{equation}
		\label{rho}
		H^2_\nu=\Big(\frac{\Dot{\nu}}{3\nu}\Big)^2=\frac{\rho_\phi}{3}\Big(1-\frac{\rho_\phi}{\bar{\rho}_\nu}\Big),\qquad\bar{\rho}_\nu=\frac{27}{4\gamma^2\beta_0^2}=\rho_\text{crit}.
	\end{equation}
Two regularizing energy densities $\bar{\rho}$ appear in the correction factors; both introduce a critical point in the evolution of $H^2$. On one hand, $\bar{\rho}_p$ depends on time $\phi$ and on the constant of motion $P_\phi$ through $p$, as we will see below, but its presence still makes it so that, when $\rho_\phi=\bar{\rho}_p$, the critical point is reached and a Big Bounce appears; on the other hand, $\bar{\rho}_\nu$ depends only on fundamental constants and is a proper critical energy density.
	
\begin{figure}
		\centering
		\includegraphics[scale=0.5]{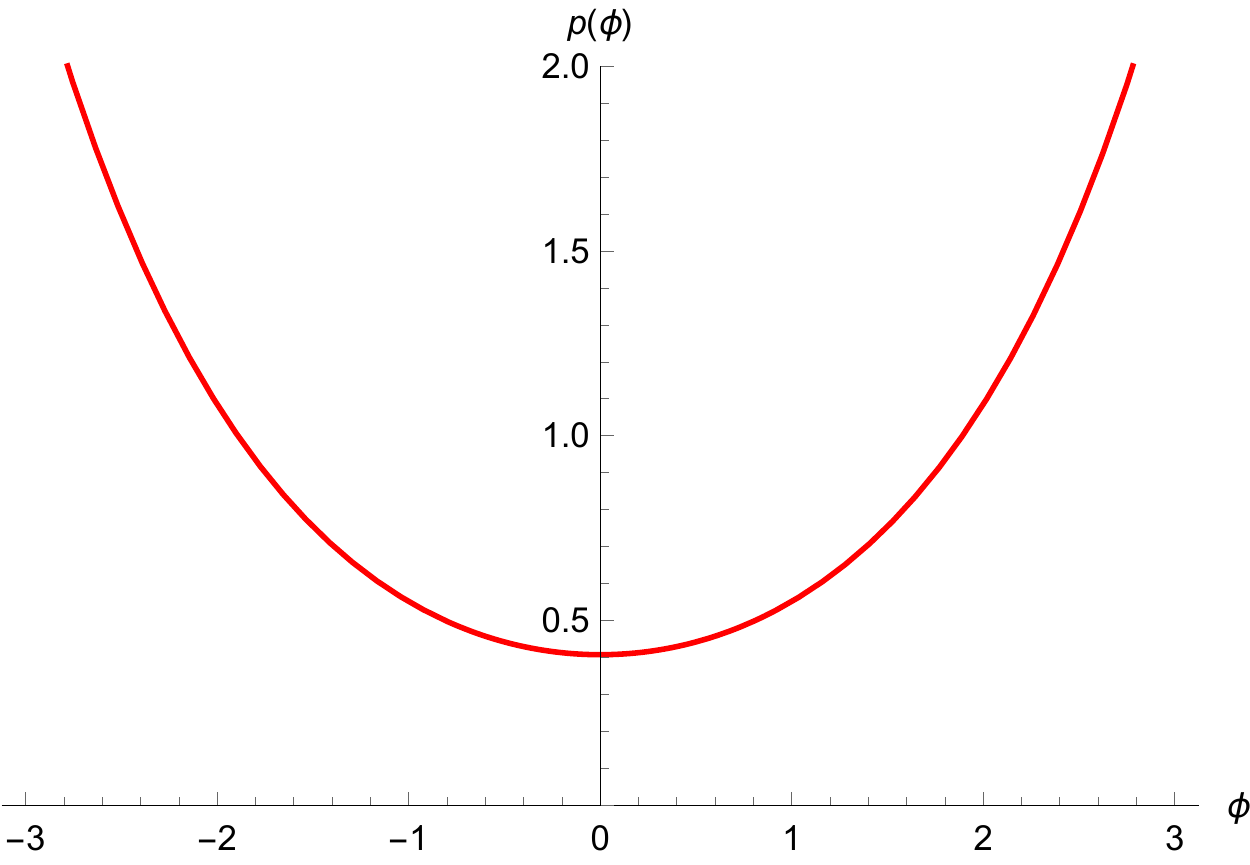}\,
		\includegraphics[scale=0.5]{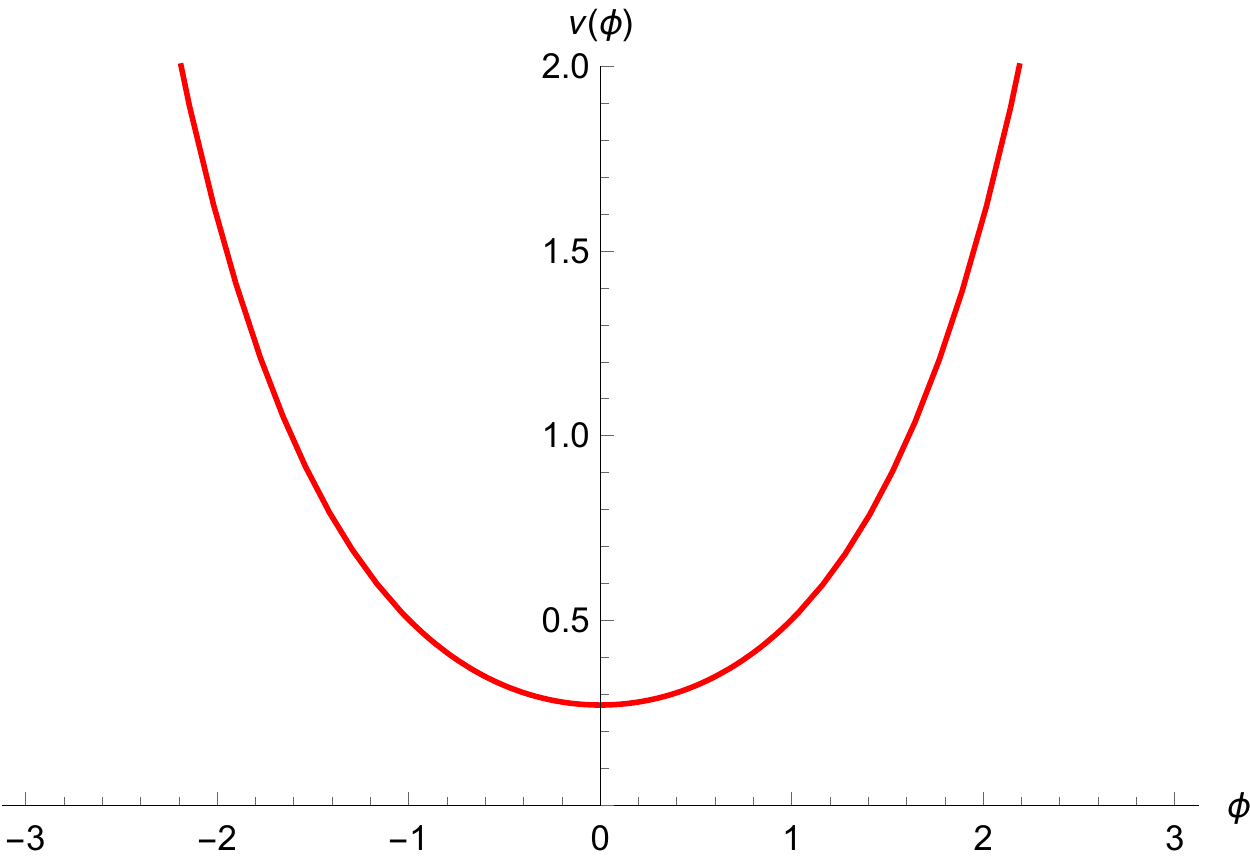}
		\caption{The polymer trajectories of $p$ (left) and $\nu$ (right) as functions of time $\phi$.}
		\label{semiclevo}
	\end{figure}

Let us now consider the scalar field $\phi$ as the internal time for the dynamics by fixing the gauge $\dot{\phi}=1$ \cite{Isham75,Giesel_2015}. The effective Friedmann equations can be solved analytically:
	\begin{equation}
		\Big(\frac{1}{p}\frac{dp}{d\phi}\Big)^2=\frac{2}{3}\Big(1-\frac{\gamma^2\beta_0^2}{6}\frac{P_\phi^2}{p^2}\Big),\qquad p(\phi)=\frac{\gamma\beta_0}{\sqrt{6}}\,P_\phi\,\cosh\left(\sqrt{\frac{2}{3}\,}\,\phi\right);
		\label{friedmannfirstcase}
	\end{equation}
	\begin{equation}
		\Big(\frac{1}{\nu}\frac{d\nu}{d\phi}\Big)^2=\frac{3}{2}\Big(1-\frac{4\gamma^2\beta_0^2}{54}\frac{P_\phi^2}{\nu^2}\Big),\qquad\nu(\phi)=\frac{2\gamma\beta_0}{3\sqrt{6}}\,P_\phi\,\cosh\left(\sqrt{\frac{3}{2}\,}\,\phi\right).
		\label{nufi}
	\end{equation}
As shown in figure \ref{semiclevo}, the polymer trajectories of $p$ and $\nu$ decrease (as expected classically) until they reach the quantum era where the effects of quantum geometry become dominant; they then reach a non-zero minimum and start to increase again. The resulting dynamics is that of a bouncing Universe replacing the classical Big Bang. For all the figures in this section we used the values $\beta_0=1/10$, $\gamma=1$ and $P_\phi=10$.
	
\subsection{Phenomenology with Particle Creation}
Here we introduce particle creation with the aim of finding a phenomenological signature of the two semiclassical schemes. Since this is a dissipative phenomenon, the assumption of constant entropy is replaced by that of constant entropy per particle.
	
By introducing a non-zero chemical potential, a new term parametrizing particle creation appears in the continuity equation for the energy density \cite{Montani:2011zz,Montani2001,CALVAO}:
	\begin{equation}
		\dot{\rho}+3H\rho(1+w)=0\quad\to\quad\dot{\rho}+3H\rho(1+w)\left(1-\abs{\dv{\ln{\mathcal{N}}}{\ln{\nu}}}\right)=0,
		\label{continuitywithparticles}
	\end{equation}
where $\mathcal{N}$ is the number of particles in the comoving volume. Given that a constant entropy per particle $s=\frac{S}{\mathcal{N}}$ is assumed, the number of particles is directly proportional to the total entropy $S$ produced. This request ensures that the entropy production is strictly due to the particle creation process, and it has no relation with the physics or the dynamics of each single particle. Furthermore, the resulting proportionality between the entropy and the particle number is immediately translated into the proportionality of the entropy density $s$ and the particle number density $n=\dv{\mathcal{N}}{\nu}$. This feature preserves a property valid in the standard cosmological picture \cite{Kolb:1990vq}, since, for instance, for the radiation component considered in the numerical analysis below (whose presence is naturally expected in the very early Universe), we have
\begin{equation}
s_\gamma=\frac{\rho_\gamma+p_\gamma}{T}=\frac{4}{3}\frac{\rho_\gamma}{T}\propto n_\gamma,
\label{ge1}
\end{equation}
where the subscript $\gamma$ refers to the radiation fluid. There is no convincing physical reason that the request of a constant entropy per particle be an assumption phenomenologically inadequate to the quantum evolution of the Universe, especially in the present scenario, in which the matter creation phenomenon is considered on an expanding polymer-modified classical background.

With the addition of this new component, the Friedmann equations \eqref{Friedmannpc} and \eqref{rho} rewrite as
\begin{equation}
	H^2_p=\frac{\rho_\phi+\rho_\gamma}{3}\Big(1-\frac{\rho_\phi+\rho_\gamma}{\bar{\rho}_p}\Big),
	\label{friedmannparticlespc}
\end{equation}
\begin{equation}
	H^2_\nu=\frac{\rho_\phi+\rho_\gamma}{3}\Big(1-\frac{\rho_\phi+\rho_\gamma}{\bar{\rho}_\nu}\Big).
	\label{friedmannparticlesnu}
\end{equation}
In order to solve this equation, we make the usual ansatz \cite{Montani:2011zz}
	\begin{equation}
		\abs{\dv{\ln{\mathcal{N}}}{\ln{\nu}}}\propto H^{2b},
		\label{ansatz}
	\end{equation}
where $b$ is a free parameter. Therefore we can rewrite the continuity equation for $\rho_\gamma(\nu)$ as
	\begin{equation}
		\dv{\rho_\gamma(\nu)}{\nu}+\frac{\rho_\gamma(\nu)}{\nu}\,(1+w)\left(1-\Big(\frac{H}{\bar{H}}\Big)^{2b}\right)=0.
		\label{contparticles}
	\end{equation}
The ansatz above has a phenomenological character and, by the direct proportionality between the particle creation rate and the Universe expansion rate $H$, the physical origin of particle creation is identified with the rapid time variation of the primordial gravitational field. In the context of a bouncing cosmology, this proportionality has the significant implication that near the Bounce, where the expansion rate $H$ vanishes, the matter creation is correspondingly suppressed. In other words, the process of matter creation has a major impact on the Universe dynamics in an intermediate region between the minimal volume and the late Universe. However, this maximum of matter creation concerns a very primordial phase, when the energy is still of comparable order to the critical value. 

The ansatz \eqref{ansatz} and the corresponding continuity equation \eqref{contparticles} contain two phenomenological parameters $b$ and $\bar{H}$. The first is taken in the numerical analysis below of order unity, since there are no reasonable indications for its deviation from the classical setting. For what concerns $\bar{H}$, the only important constraint comes from avoiding that matter creation affects the de-Sitter phase of an inflationary scenario. Otherwise, in the opposite case, we could obtain a spectrum of inhomogeneous perturbations that is different than the natural scale invariant one \cite{Weinberg:1972kfs}. This consideration suggests that the value of $\bar{H}$ must be such that the matter creation is strongly suppressed before inflation starts, for example for a Universe temperature of order $\lesssim10^{15}\,GeV$. However, in our study the value of $\bar{H}$ is chosen to obtain a maximum of the matter creation nearby the Planckian phase of the Universe (where the polymer modifications are relevant), as stressed above. Finally, we consider the contribution of all relativistic species as a single effect, and therefore the value of $\bar{H}$ is actually meant to represent an average effect over all ultrarelativistic matter species.

Note that modifying the continuity equation as in \eqref{continuitywithparticles} while keeping the same form for the modified Friedmann equations (except for the presence of the new radiation component), unavoidably leads to a modified form of the acceleration equations as follows:
\begin{equation}
    \frac{\ddot{p}}{2p}=\frac{\rho}{2}\left(\frac{1}{3}-w(1-2\frac{\rho}{\bar{\rho}_p})+\abs{\dv{\ln{\mathcal{N}}}{\ln{\nu}}}(1+w)(1-2\frac{\rho}{\bar{\rho}_p})\right),
\end{equation}
\begin{equation}
    \frac{\ddot{\nu}}{3\nu}=\frac{\rho}{2}\left(1-w(1-2\frac{\rho}{\bar{\rho}_\nu})+\abs{\dv{\ln{\mathcal{N}}}{\ln{\nu}}}(1+w)(1-2\frac{\rho}{\bar{\rho}_\nu})\right).
\end{equation}
The new term $\abs{\dv{\ln{\mathcal{N}}}{\ln{\nu}}}$ acts as a positive energy density and could therefore drive the acceleration when $\rho<\frac{\bar{\rho}}{2}$, that is away from the Planck regime; however, given the ansatz \eqref{ansatz}, in the late universe the Hubble parameter decays and particle creation is strongly suppressed, so the additional term cannot act as a suitable candidate to explain a de-Sitter-like phase of expansion, as suggested for example in \cite{Particles89,LimaParticles,Balfagon_2015}.

\begin{figure}
		\centering
		\includegraphics[width=0.48\linewidth]{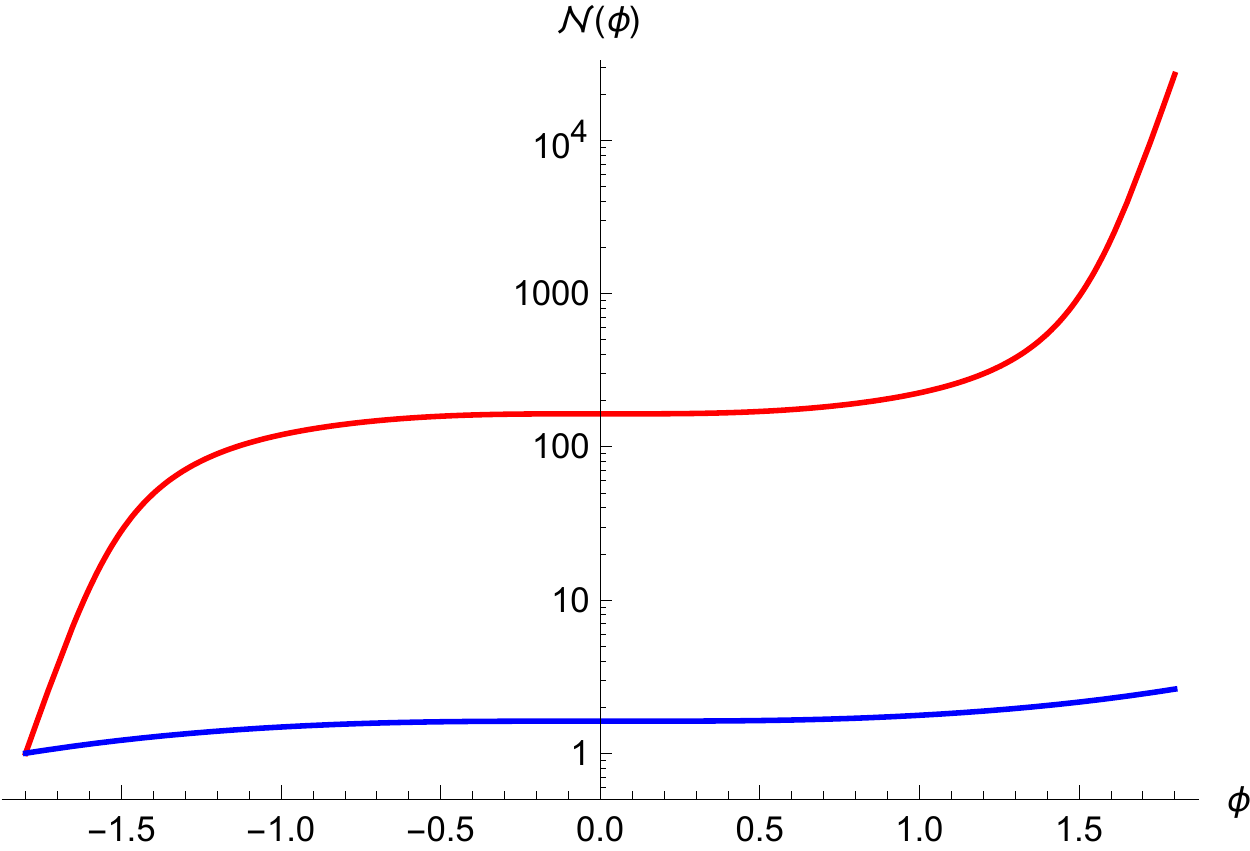}
		\,
		\includegraphics[width=0.48\linewidth]{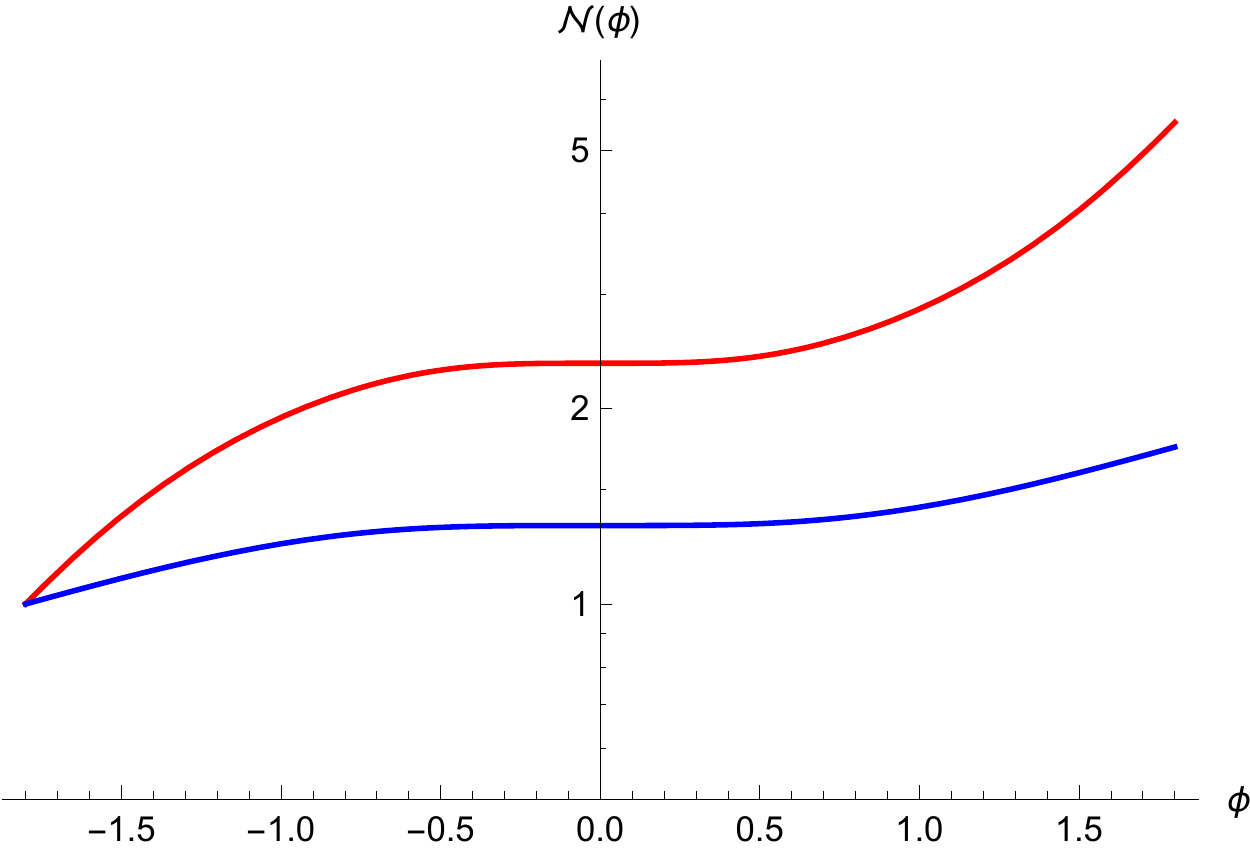}
		\caption{Number of particles as function of time $\phi$ (Ashtekar variables in blue, volume variables in red) for two different values of the initial condition ($P_\phi$=4.5 on the left, $P_\phi=5.5$ on the right).}
		\label{number}
\end{figure}

\begin{figure}
		\centering
		\includegraphics[width=0.48\linewidth]{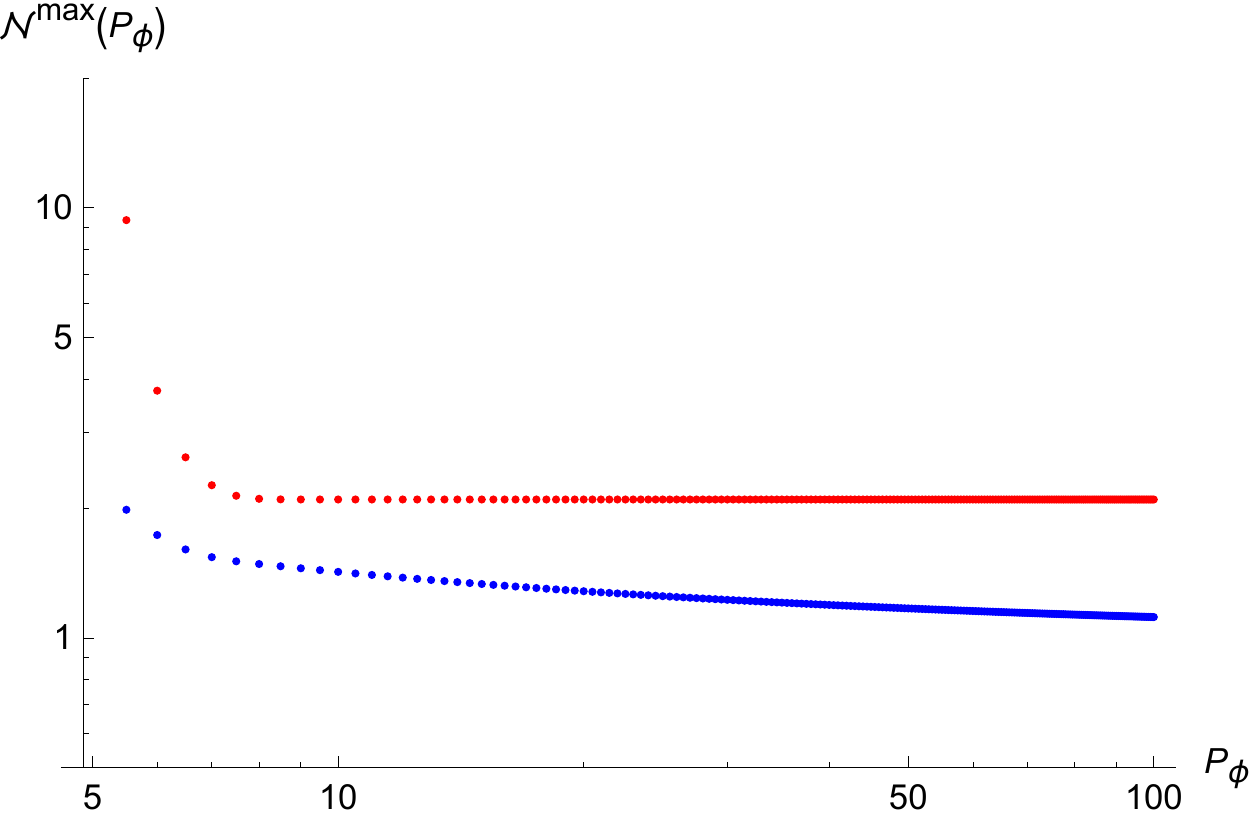}
		\,
		\includegraphics[width=0.48\linewidth]{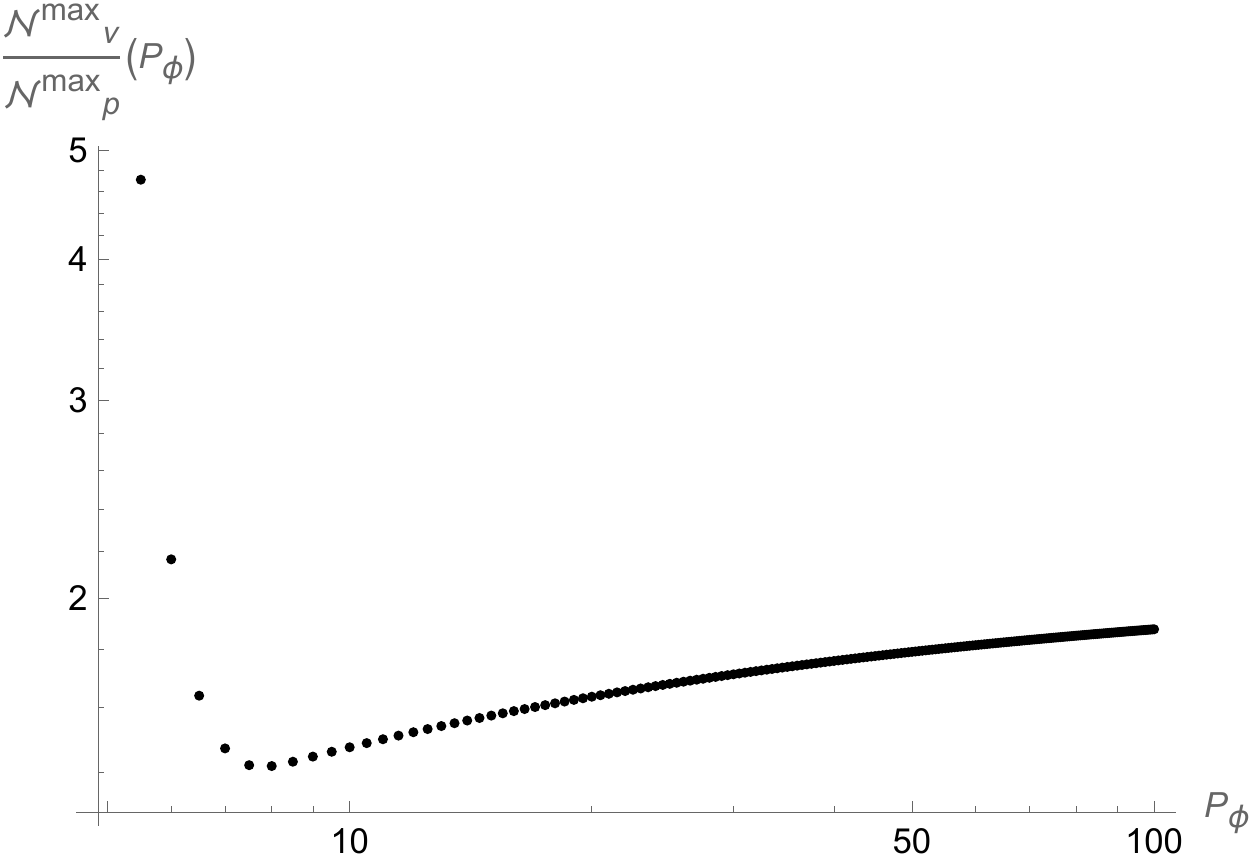}
		\caption{Left: number of final particles as function of the initial condition $P_\phi$; blue and red dots refer respectively to the Ashtekar variables and the volume variables. Right: ratio between the numbers of final particles in the two cases.}
		\label{finalparticles}
\end{figure}

Now it is possible to numerically solve eq. \eqref{contparticles} for $\rho_\gamma(\nu)$ and to find $\nu(\phi)$ and $\mathcal{N}(\phi)$ from the Friedmann equations \eqref{friedmannparticlespc}, \eqref{friedmannparticlesnu} and from the ansatz \eqref{ansatz} respectively. In figg. \ref{number} we see the evolution of the total number of particles as function of time $\phi$ for two different values of the initial condition $P_\phi$; as we can see, in the case of the Ashtekar variables the production of particles (and therefore of entropy) is negligible with respect to the volume representation. In figg. \ref{finalparticles} we see the final number of particles in the two cases and the ratio between them as function of the initial condition $P_\phi$; we see how a greater value of $P_\phi$ (and therefore a more dominant scalar field density) suppresses the creation of particles in both cases, and the ratio reaches an asymptotic value, while on the other hand for small values of $P_\phi$ the ratio grows appreciably. It is worth stressing that the different behaviour outlined above between the entropy creation in the polymerization of the Ashtekar-Barbero-Immirzi connection or of the volume variable has a precise physical meaning in the polymer paradigm only. However, since in the considered semiclassical dynamics when matter creation is absent the two polymer pictures mimic the $\mu_0$ and $\bar{\mu}$ schemes of LQC respectively, we are legitimated to suppose that our results might be valid also in the LQC theory and could represent an important phenomenological difference between the two schemes.
	
	
\section{Polymer quantum dynamics of the FLRW Universe}
\label{quantum}
In this section the main purpose is to promote the system to a quantum level, starting from the Hamiltonian constraint in its quantum counterpart and applying Dirac quantization \cite{matschull1996diracs} directly to quantum wavefunctions in order to obtain the WDW equation. The variables are directly promoted to a quantum level, the Poisson brackets to commutators and the constraints to operators; the latter, when applied to the quantum states, will select physical states and yield the WDW equation $\hat{\mathcal{C}}\ket{\Psi}=0$. This procedure will lead to the dynamics whereby the system will fix $\Psi$ as an eigenstate for the Hamiltonian with vanishing eigenvalue.
	
\subsection{Quantum analysis in the Ashtekar variables}
To implement Dirac quantization method we promote variables to quantum operators as
	\begin{equation}
		\hat{p}=-i\frac{\gamma}{3}\frac{d}{dc},\qquad\hat{c}=\frac{1}{\beta_0}\mbox{sin}(\beta_0c),\qquad\hat{P}_\phi=-i\frac{d}{d\phi}.
		\label{operatorspc}
	\end{equation}
Given the modified Hamiltonian \eqref{Cpoly}, the Hamiltonian constraint operator in the momentum representation is
	\begin{equation}
		\hat{\mathcal{C}}_\text{poly}\Psi(c,\phi)=\Big[-\frac{2}{3\beta_0^2}\Big(\mbox{sin}(\beta_0 c)\frac{d}{dc}\Big)^2+\frac{d^2}{d\phi^2}\Big]\Psi(c,\phi)=0.
		\label{wheelerpoly}
	\end{equation}
This mixed factor ordering allows us to interpret this differential equation as a Klein-Gordon-like equation; indeed it admits a conserved current and we can write the scalar product as
\begin{equation}
    \ev{\hat{O}}{\Psi}=\int_{-\frac{\pi}{\beta_0}}^{\frac{\pi}{\beta_0}}\frac{dc}{\frac{\sqrt{2}}{\beta_0\sqrt{3}}\sin(\beta_0c)}\,\,i\big(\Psi^*\,\partial_\phi(\hat{O}\Psi)-(\hat{O}\Psi)\,\partial_\phi\Psi^*\big).
    \label{KGscalarproductc}
\end{equation}
Thanks to the substitution to the auxiliary variable $x=\sqrt{\frac{3}{2}\,}\,\ln{\abs{\tan\Big(\frac{\beta_0 c}{2}\Big)}}+x_0$, equation \eqref{wheelerpoly} becomes a massless Klein-Gordon equation:
\begin{equation}
	\label{KG}
	\frac{d^2}{dx^2}\Psi(x,\phi)=\frac{d^2}{d\phi^2}\Psi(x,\phi),
\end{equation}
where $\Psi$ can be written as a planewave superposition: $\Psi(x,\phi)=\chi(x)e^{-ik_\phi\phi}$. The solution to this equation can be stated in the form of a Gaussian-like localized wavepacket:
	\begin{equation}
		\Psi(x,\phi)=\int_0^{\infty}dk_\phi\,\frac{e^{-\frac{(k_\phi-\overline{k}_\phi)^2}{2\sigma^2}}}{\sqrt{4\pi\sigma^2}}\,k_\phi\,e^{ik_\phi x}e^{-ik_\phi\phi}.
	\end{equation}
Here $k_\phi$ is the energy-like eigenvalue of the operator $\hat{P}_\phi$, and its positive or negative values select collapsing or expanding solutions respectively.
	
Now, in order to investigate the non-singular behaviour of the model, we can compute the expectation value of the energy density operator $\hat{\rho}_\phi=\frac{\hat{P}_\phi^2}{2\hat{p}^3}$. In what follows, all the mean values and variances of the relevant operators will be computed using the auxiliary variable $x$ inside the scalar product \eqref{KGscalarproductc}, so that the results are given by
\begin{equation}
		\ev{\hat{O}}{\Psi}=\int_{-\infty}^{\infty}dx\,\,i\big(\Psi^*\,\partial_\phi(\hat{O}\Psi)-(\hat{O}\Psi)\,\partial_\phi\Psi^*\big),
		\label{KGev}
	\end{equation}
where we assumed normalized wavefunctions. We remark that the transformation from $c$ to $x$ in the minisuperspace preserves the expectation values of the physical observables.

\begin{figure}
		\centering
		\includegraphics[scale=0.5]{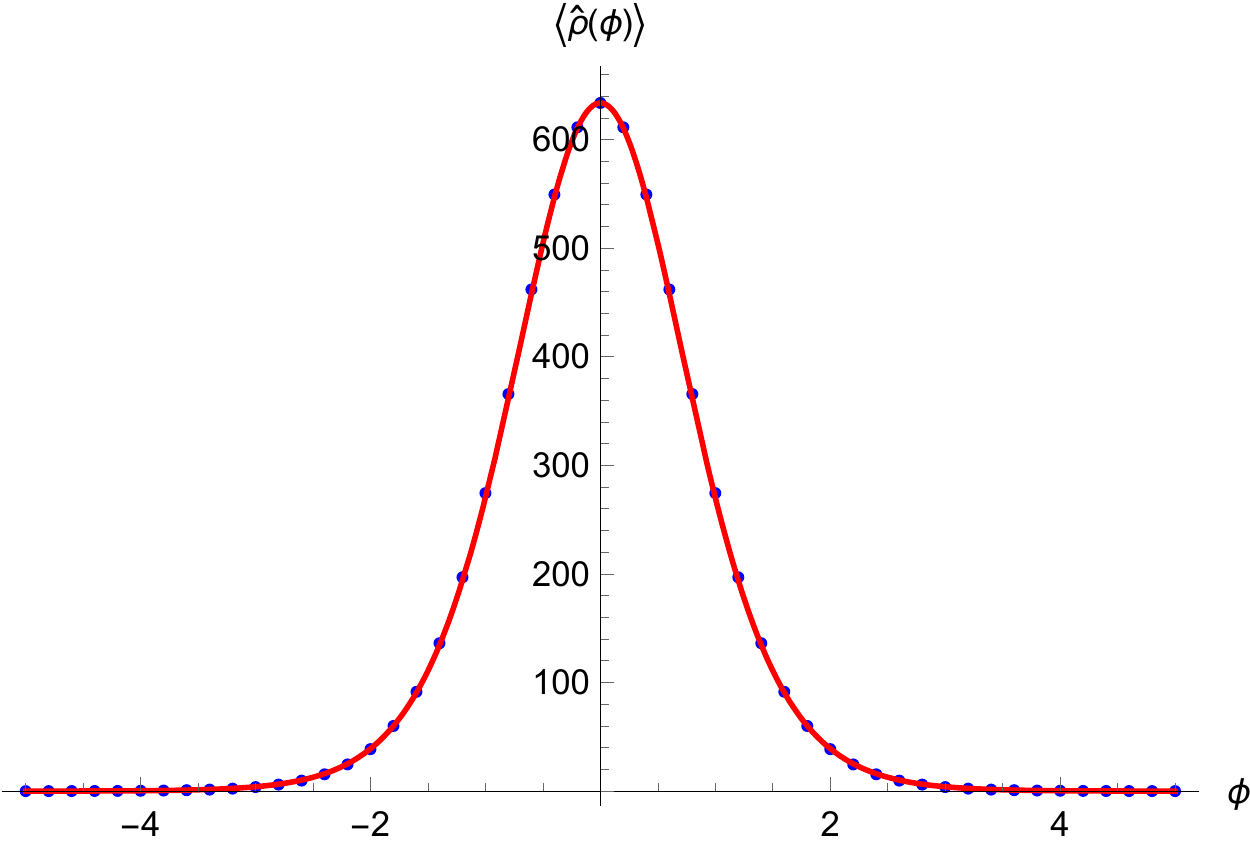}
		\,
		\includegraphics[scale=0.5]{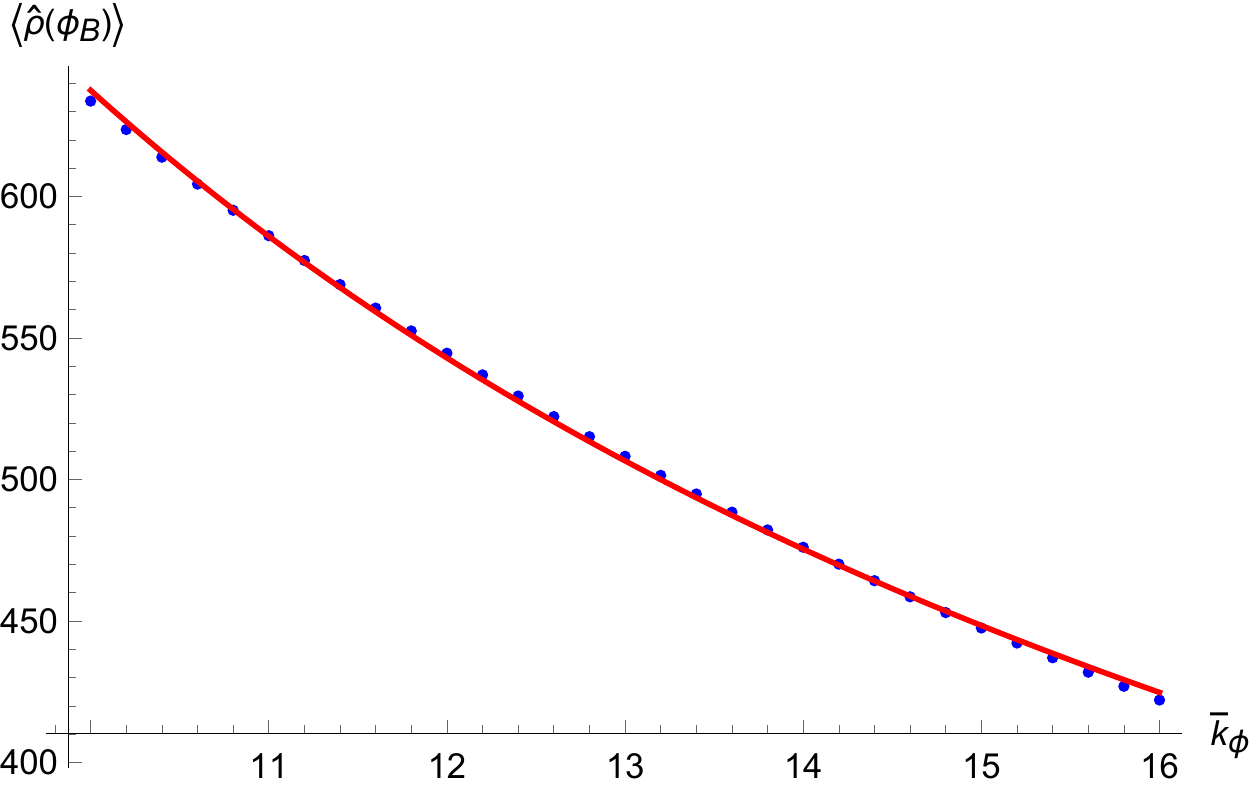}
		\caption{Left: the expectation value of the energy density as function of time for $\overline{k}_\phi=10$ (blue dots); right: the expectation value of the energy density at the time $\phi_B$ of the Bounce as function of $\overline{k}_\phi$ (blue dots). Both have been fitted with a function in accordance with the semiclassical evolution (full red line).}
		\label{rho(phi)rhomaxpc}
	\end{figure}
	
In figure \ref{rho(phi)rhomaxpc} we show the time dependence of $\ev{\hat{\rho}_\phi(\phi)}$ for a fixed value of $\overline{k}_\phi$ and the maximum $\ev{\hat{\rho}_\phi(\phi_B)}$, i.e. the expectation value of the Bounce density, that results to be inversely proportional to $\overline{k}_\phi$, in accordance with the semiclassical critical energy density \eqref{Friedmannpc}. The points representing the quantum expectation values were obtained through numerical integration and have been fitted with the full lines; they are in accordance with the semiclassical trajectories when taking into account numerical effects and quantum fluctuations. Note that the action of the energy density operator expressed as function of $x$ has been simplified thanks to the hypothesis of a sufficiently localized wavepacket. For the figures in this section we used the values $\sigma=3.5$, $\beta_0=1/10$, $\gamma=1$ and, when needing a fixed value, $\overline{k}_\phi=10$.
	
The non-diverging nature of the energy density of the primordial Universe clearly implies, in view of its scalar and physical nature, the existence of a minimum non-zero volume (although strongly dependent on the initial conditions), thus confirming the replacement of the singularity with a Bounce also in the quantum system. A more precise assessment of the nature of the Bounce would require also a careful analysis of the variance of the density and the moments of the quantum probability, as done for instance in \cite{BojowaldMoments}, but already at this level the existence of a Bounce is a solid prediction of our model.
	
\subsection{Quantum analysis in the volume variable}
We now implement the quantization procedure on the system expressed in the volume variable $\nu$ and its conjugate momentum $\tilde{c}$. This procedure is very similar to the previous one and to the equivalent Loop quantization performed in \cite{Ashtekar2_2006,Ashtekar_2011review,Ashtekar_2008}. Therefore we will again report only the most relevant results.
	
The fundamental variables, when promoted to operators, act as
	\begin{equation}
		\hat{\nu}=-i\frac{\gamma}{3}\frac{d}{d\tilde{c}},\qquad\hat{\tilde{c}}=\frac{1}{\beta_0}\mbox{sin}(\beta_0 \tilde{c}),\qquad\hat{P}_\phi=-i\frac{d}{d\phi},
		\label{operatorsVc}
	\end{equation}
and the quantum Hamiltonian constraint becomes
	\begin{equation}
		\hat{\tilde{\mathcal{C}}}_{\text{poly}}\Psi(\Tilde{c},\phi)=\Big[-\frac{3}{2\beta_0^2}\Big(\mbox{sin}(\beta_0 \Tilde{c})\frac{d}{d\Tilde{c}}\Big)^2+\frac{d^2}{d\phi^2}\Big]\Psi(\Tilde{c},\phi)=0.
		\label{wheelerpoly2}
	\end{equation}
In this case the appropriate substitution to use is simply $\tilde x=\sqrt{\frac{2}{3}\,}\ln{\abs{\tan\Big(\frac{\beta_0 \Tilde{c}}{2}\Big)}}+\tilde x_0$; this leads us to the same massless Klein-Gordon equation of the previous case and allows us to write its solution in the $\Tilde{x}$-representation as a Gaussian-like wavepacket in the form
	\begin{equation}
		\Psi(\Tilde{x},\phi)=\int_0^{\infty}dk_\phi\hspace{4pt}\frac{e^{-\frac{(k_\phi-\overline{k}_\phi)^2}{2\sigma^2}}}{\sqrt{4\pi\sigma^2}}\,k_\phi\,e^{ik_\phi\tilde x}e^{-ik_\phi\phi}.
	\end{equation}
	
\begin{figure}[htbp]
		\centering
		\includegraphics[scale=0.5]{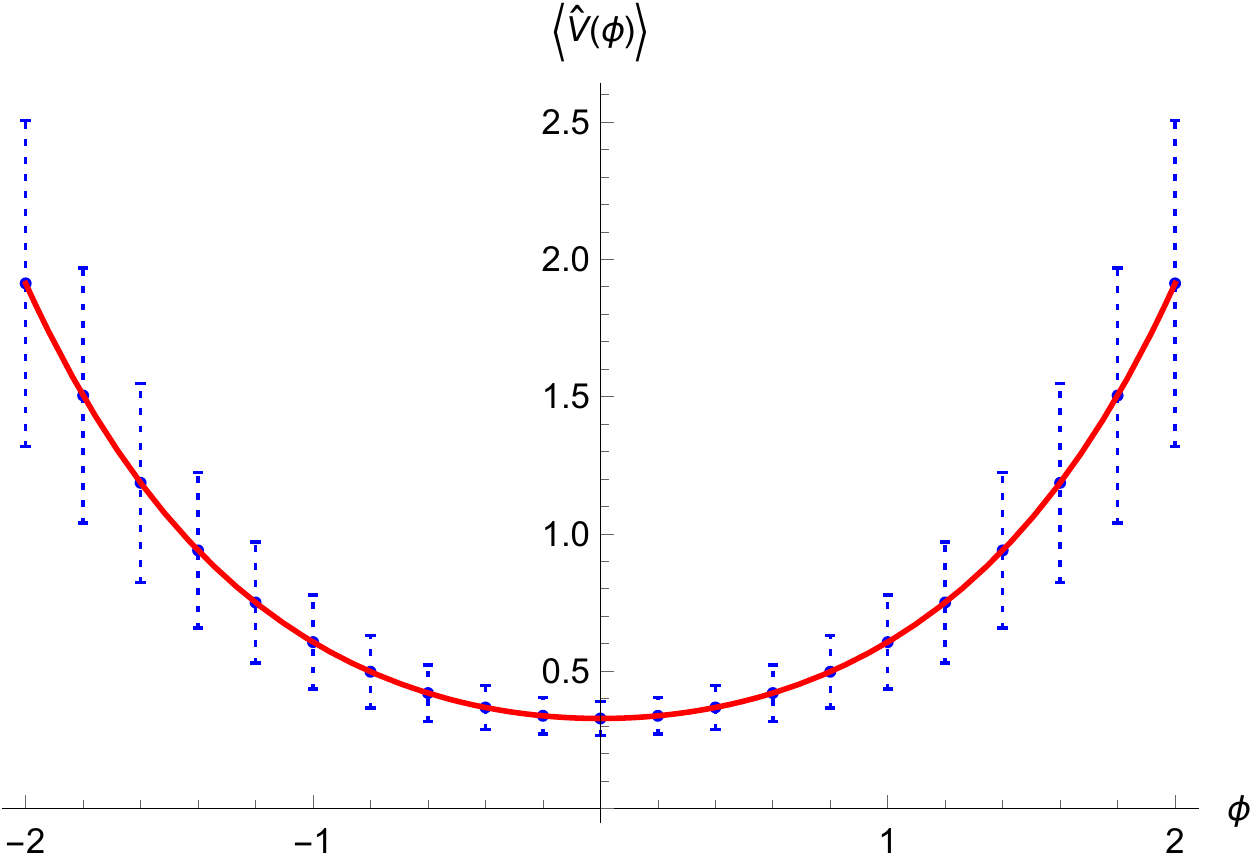}
		\,
		\includegraphics[scale=0.5]{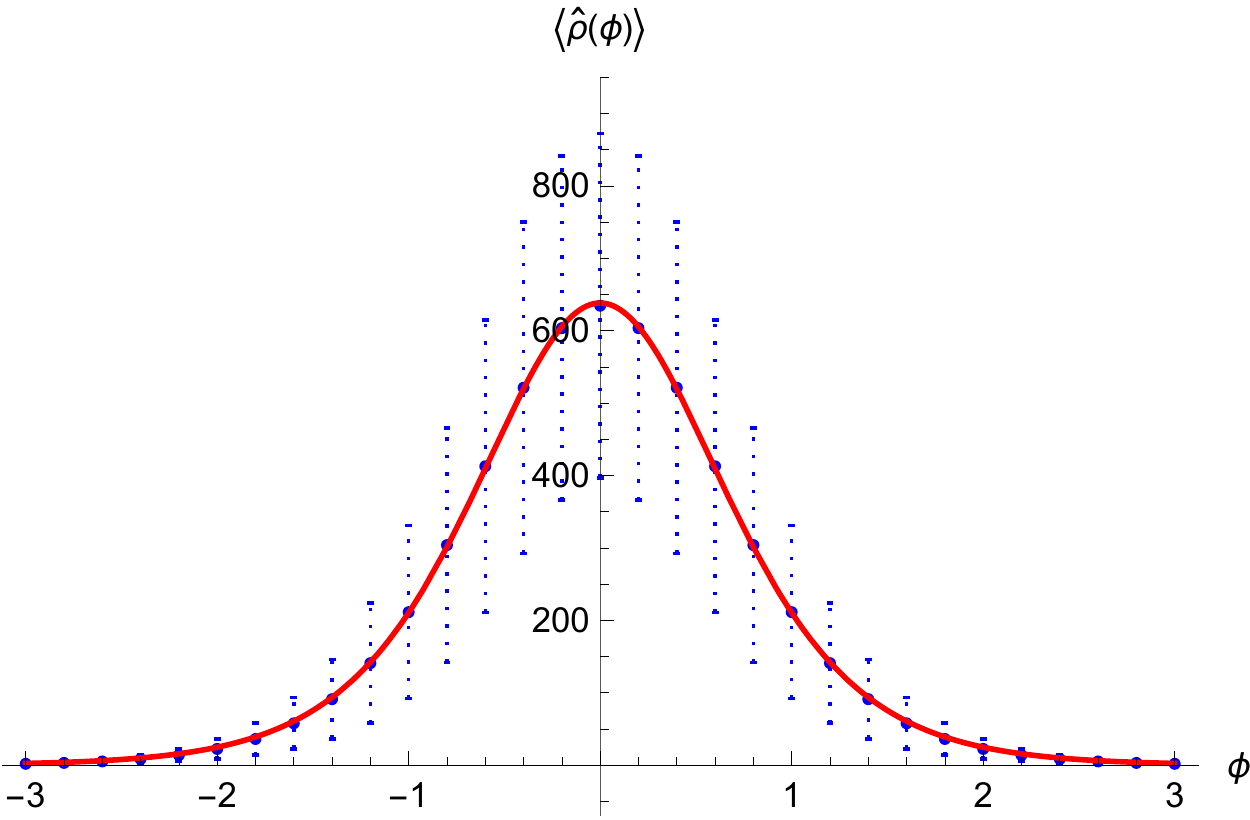}
		\caption{Expectation values of the volume (left) and of the energy density (right) as functions of time for $\overline{k}_\phi=10$ (blue dots), fitted with functions in accordance with the semiclassical evolution (full red lines). The error bars are calculated as the standard deviation of the corresponding operator.}
		\label{V(phi)RhoVc}
	\end{figure}

\begin{figure}
		\centering
		\includegraphics[scale=0.5]{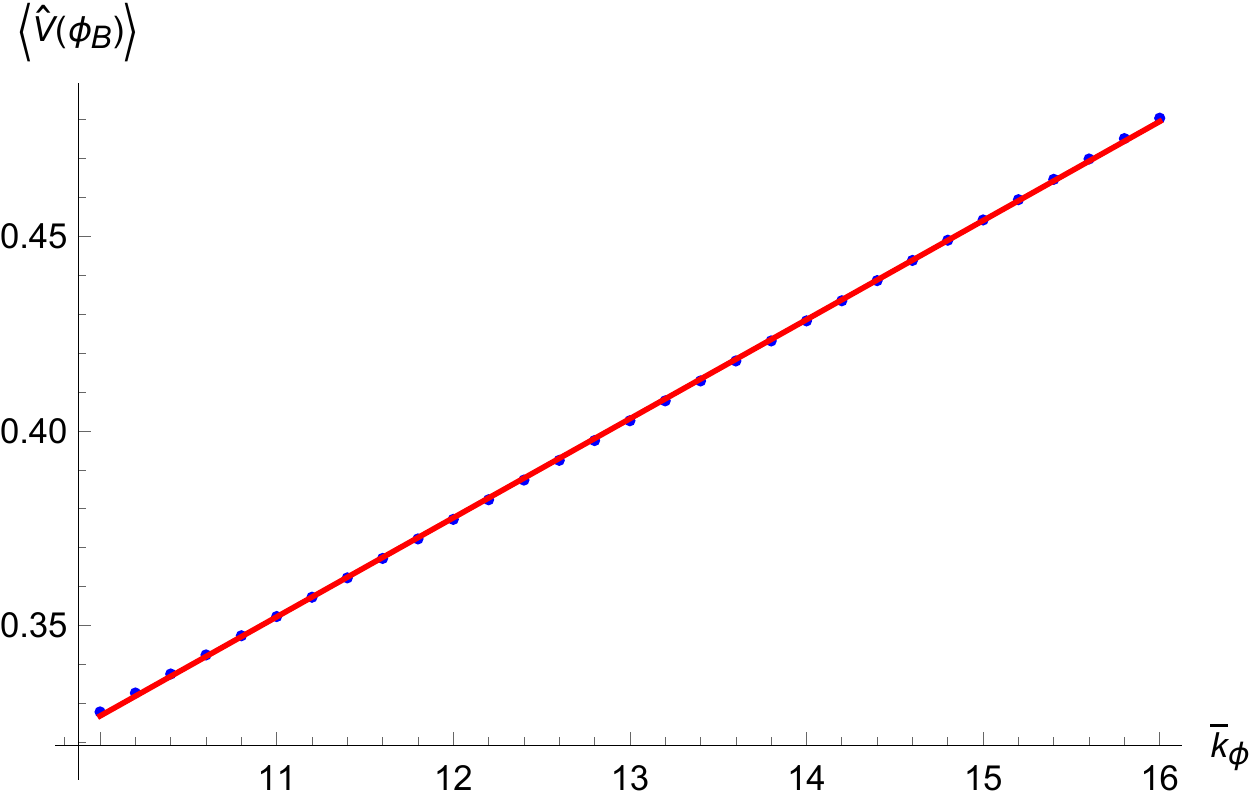}
		\,
		\includegraphics[scale=0.5]{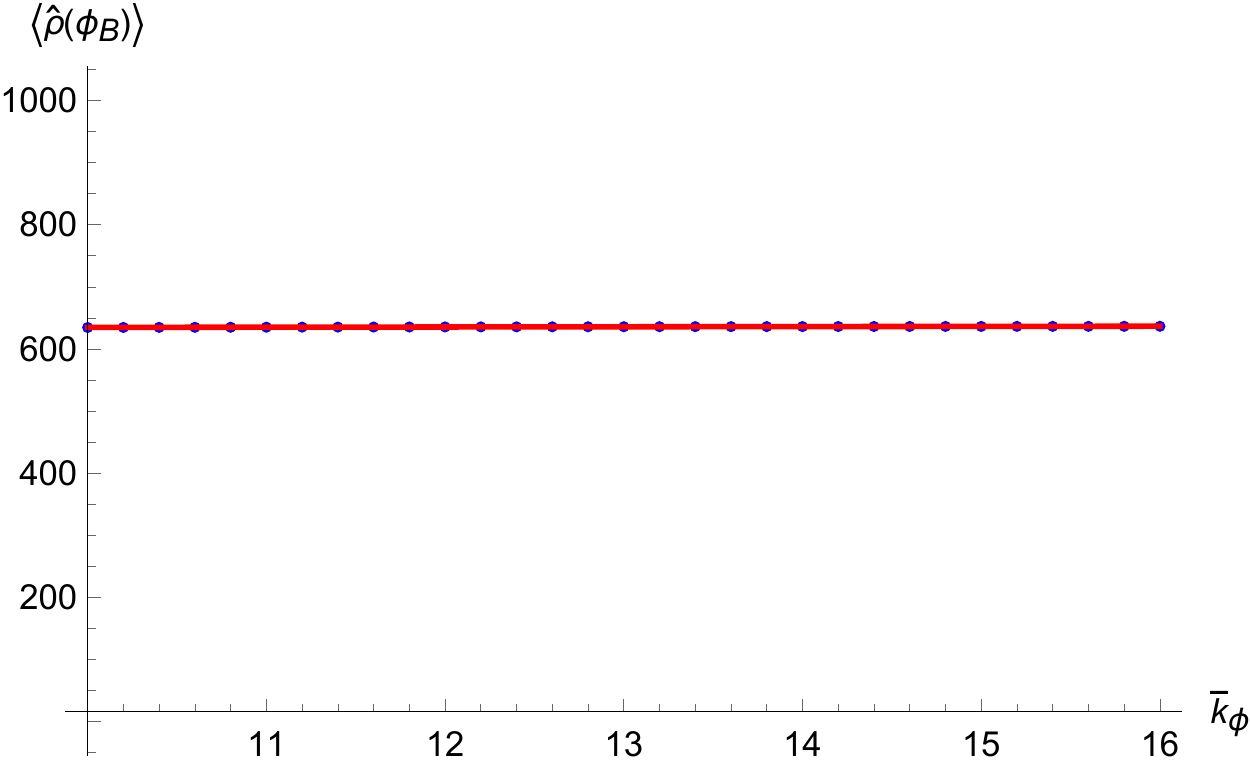}
		\caption{Expectation values of the volume (left) and of the density (right) at the time $\phi_B$ of the Bounce as functions of $\overline{k}_\phi$ (blue dots), fitted with functions in accordance with the semiclassical evolution (full red lines).}
		\label{VminRhomaxVc}
	\end{figure}

Now the procedure is exactly the same as before; the operators of interest in this case are both the volume and the energy density, that in this representation act as $\hat{V}=\hat{\nu}$ and $\hat{\rho}_\phi=\frac{\hat{P}_\phi^2}{2\hat{\nu}^2}$. Again we use the same Klein-Gordon scalar product \eqref{KGscalarproductc} (except for a numerical constant) to calculate their expectation value; their action is derived from \eqref{operatorsVc}. In figures \ref{V(phi)RhoVc} we can see the expectation values of the volume $\langle\hat{V}(\phi)\rangle$ and density $\ev{\hat{\rho}_\phi(\phi)}$ as functions of time. In the left panel of figure \ref{VminRhomaxVc} the value $\langle\hat{V}(\phi_B)\rangle$ of the volume at the Bounce is shown as function of the initial value $\overline{k}_\phi$; the minimum volume scales linearly with the energy-like eigenvalue, in accordance with the semiclassical expression for $\nu(\phi)$ given in \eqref{nufi}. Then in the right panel of the same figure we see the Bounce density $\ev{\hat{\rho}_\phi(\phi_B)}$ for different values of $\overline{k}_\phi$; in accordance with the semiclassical critical energy density \eqref{rho}, the density at the Bounce in the new variables does not depend on the initial conditions of the system (or of the wavepacket in the quantum analysis). For the parameters we used the same values as before ($\sigma=3.5$, $\beta_0=1/10$, $\gamma=1$ and $\overline{k}_\phi=10$ when necessary).
	
The fact that in this set of variables the critical energy density of the Universe is fixed and does not depend on initial conditions is related to the volume itself being chosen as the configurational variable for the polymer quantization of the system. However, even if this representation is preferable on the grounds that it yields an universal Bounce scenario that is physically more acceptable, this choice is clearly dynamically inequivalent to the Ashtekar variables $(p,c)$ that are the only $SU(2)$ choice in LQC. Actually, in the next section we will show that, although it is possible on a semiclassical level to recover the physical equivalence between the two sets of variables, this leads in the polymer representation to a translational operator whose implementation on the states constitutes a non-trivial issue.
	
\section{Discussion of the results}\label{discussion}
Above we stressed that in the polymer framework the Universe always possesses a bouncing point in the past both in a semiclassical and in a pure quantum description, with the difference that when the natural connection $c$ is used the maximal density is fixed by the initial conditions on the system, while when using the redefined variable $\nu$ the Bounce density depends on fundamental constants and the Immirzi parameter only. In this respect, we observe that the polymer quantization introduces a minimal value to the geometrical operators area and volume when $p\propto a^2$ and $\nu\propto a^3$ respectively. In the first case, by discretizing the area element also the volume results to be regularized since a bouncing cosmology emerges, but with different implications on the behaviour of the critical energy density. Consequently, these two representations clearly appear dynamically and physically not equivalent (see \cite{Antonini_2019,Montani_2019,Crin__2018,Giovannetti_2019} for similar not equivalent behaviours in polymer cosmology). However, we want to stress that both the polymerization procedures, in terms of the Ashtekar-Barbero-Immirzi connection and the volume variable, have some physical link to the background LQC kinematics. The first is justified by the direct interpretation in terms of the right $SU(2)$ connection adopted in LQC, while the latter refers to the kinematical result about its spectrum discretization \cite{Rovelli_1995DiscretenessAreaVolume}. In other words, polymerizing the Ashtekar-Barbero-Immirzi connection means giving LQG features to the natural variable in which the Loop setup is formulated; on the same footing, polymerizing the volume corresponds to attributing a discrete structure to the quantum representation of this geometrical operator, as in the original LQG theory.
	
\subsection{Linking the Two Pictures}
Let us now compare the semiclassical dynamics in both sets of variables in search of a physical link between the two representations. If we start with the polymer-modified system in the $(\nu,\tilde{c})$ representation, the canonical transformation to the natural Ashtekar connection is
	\begin{equation}
		\label{transf}
		p=\nu^\frac{2}{3}\qquad c=\frac{3}{2}\Tilde{c}\nu^\frac{1}{3}.
	\end{equation}
However, to realize a canonical transformation in the polymer construction, we have to introduce the condition $\beta_0\Tilde{c}=\beta_0'c$ in order to map the polymer Hamiltonian written in the variables $(\nu,\Tilde{c})$ to that one \eqref{Cpoly} written in the new variables $(p,c)$ and make the polymer-modified Poisson brackets formally invariant: $\pb{\Tilde{c}}{\nu}=\frac{\gamma}{3}\sqrt{1-(\beta_0\tilde c)^2}=\frac{\gamma}{3}\sqrt{1-(\beta_0'c)^2}=\pb{c}{p}$, where the substitution \eqref{ppoly} has been used on both $c$ and $\tilde{c}$. In other words, we have to deal with a new polymer parameter that depends on the configurational variable as follows \cite{Montani_2019}:
	\begin{equation}
		\label{beta'}
		\beta_0'=\frac{2}{3}\beta_0\nu^{-\frac{1}{3}}=\frac{2}{3}\frac{\beta_0}{\sqrt{p}}.
	\end{equation}
The last dependence of $\beta_0'$ on $p$ is the same of $\bar{\mu}(p)$ in the $\bar{\mu}$ scheme of LQC \cite{Ashtekar2_2006,Ashtekar_2011review,Ashtekar_2008}.
	
After introducing a dependence of the polymer parameter from the configurational variable under a canonical transformation, it is commutative to write the transformed Hamiltonian and to introduce the polymer substitution \eqref{p2sin2}. Therefore we expect that also the equations of motion for the two different sets of variables will be mapped using \eqref{transf} and \eqref{beta'}: we have
	\begin{subequations}
		\begin{equation}
			%\begin{aligned}
			\dot{p}=\frac{2}{3}\nu^{-\frac{1}{3}}\dot{\nu}=\frac{\text{N}\gamma}{3}\frac{\partial\mathcal{C}_{\text{poly}}(c,p)}{\partial c}=-\frac{2\text{N}}{\gamma\beta_0'}\sqrt{p}\sin(\beta_0'c)\cos(\beta_0'c),
			\label{p}
			%\end{aligned}
		\end{equation}
		\begin{equation}
			%\begin{aligned}
			\dot{c}=\frac{3}{2}\dot{\Tilde{c}}\nu^\frac{1}{3}+\frac{1}{2}\Tilde{c}\nu^{-\frac{2}{3}}\dot{\nu}=-\frac{\text{N}\gamma}{3}\frac{\partial\mathcal{C}_{\text{poly}}(c,p)}{\partial p}=\frac{\text{N}}{\sqrt{p}}\bigg(\frac{\sin^2(\beta_0'c)}{2\gamma\beta_0'^2}-\frac{c}{\gamma\beta_0'}\sin(\beta_0'c)\cos(\beta_0'c)+\frac{\gamma P_\phi^2}{4p^2}\bigg).
			\label{ci}
			%\end{aligned}
		\end{equation}
	\end{subequations}
For comparison, the equations of motion in the $(p,c)$ representation are
	\begin{subequations}
		\begin{equation}
			\label{pi}
			\dot p=-\frac{2\mbox{N}}{\gamma\beta_0}\sqrt{p}\hspace{4pt}\mbox{sin}(\beta_0 c)\mbox{cos}(\beta_0 c),
		\end{equation}
		\begin{equation}
			\label{c}
			\dot c=\frac{\text{N}}{3}\Big(\frac{3}{\gamma\beta_0^2}\frac{1}{2\sqrt{p}}\mbox{sin}^2(\beta_0 c)+\frac{3\gamma}{4}\frac{P_\phi^2}{p^{5/2}}\Big).
		\end{equation}
	\end{subequations}
We note that \eqref{p} is formally the same as \eqref{pi}, but the relation $\beta_0'=\frac{2}{3}\beta_0\frac{1}{\sqrt{p}}$ changes the solution, while \eqref{ci} for the connection $c$ results to be different from \eqref{c} because of the dependence of the polymer parameter $\beta_0'$ on $p$. Therefore, on a semiclassical level, there exists a physical equivalence in the evolution of $p$ and $\nu$. Indeed, thanks to \eqref{beta'}, the regularizing density $\bar{\rho}_p$ in \eqref{Friedmannpc} turns out to be the same critical energy density $\rho_\text{crit}$ of \eqref{rho}:
	\begin{equation}
		\bar{\rho_p}=\frac{3}{{\gamma^2\beta_0'}^2p}=\frac{27}{4{\gamma^2\beta_0}^2}=\rho_{\text{crit}}\,.
	\end{equation}
Also the effective Friedmann equation in the time gauge $\dot{\phi}=1$ reads as
	\begin{equation}
		\Big(\frac{1}{p}\frac{dp}{d\phi}\Big)^2=\frac{2}{3}\Big(1-\frac{4\gamma^2\beta_0^2}{54}\frac{P_\phi^2}{p^3}\Big)
	\end{equation}
and it clearly reduces to \eqref{nufi} using \eqref{transf}. The Bounce of the Universe volume has the same properties in the two sets of variables only if we consider the polymer parameter $\beta_0'$ to be dependent on $p$.
	
However, in its natural formulation PQM is associated to a lattice (that has a constant spacing by construction) only after the dynamics is assigned and after the change of variables in the classical Hamiltonian has been performed, and indeed the difference between the two schemes consists in choosing the variable for which the lattice parameter is constant: if we transform into the natural Ashtekar connection from the volume-like momentum after the polymer framework has been implemented, we have to deal with a polymer parameter depending on the configurational coordinate, and unfortunately this request prevents a full quantum analysis of the problem since it produces a translational operator that cannot be implemented. This analysis highlights the privileged nature of the variable for which the polymer parameter is taken constant, since the physical results depend on it. In particular, if in polymer semiclassical cosmology one starts from assigning a lattice in the Ashtekar variables and then canonically transforms to the volume ones, obtaining a non-constant spacing, the resulting cosmology would still be a bouncing one whose cut-off energy density depends on the initial conditions; on the other hand, starting from the volume variables, one would have a universal bouncing cosmology in both sets, with the difference that the polymer parameter would not be constant in the Ashtekar ones. Hence in PQC the use of the Ashtekar or volume variables leads to two different physical pictures. Far from defining an equivalence between the $\mu_0$ and the $\bar{\mu}$ schemes, we simply stress that the latter would just correspond to dealing with a polymer parameter depending on the configurational variable when seen in the Ashtekar-Barbero-Immirzi connection instead of the volume one. This result encourages the thought that the conclusions gained in \cite{Ashtekar_2006} after a change of basis has been performed can be obtained on the semiclassical level (and hopefully in the quantum too) also in the Ashtekar variables: the critical density of the Universe takes an absolute value, independent on the initial conditions on the semiclassical system or on the quantum wave packet.
	
We conclude by observing that the question we are addressing has a deep physical meaning since it involves the real nature of the so-called Big Bounce: is it an intrinsic cut-off on the cosmological dynamics or is it a primordial turning point fixed by initial conditions on the quantum Universe? The present analysis suggests that the second case appears more natural in PQC if it is referred to LQG, since the quantum implementation of the Ashtekar connection produces results in accordance with the original analysis in \cite{Ashtekar_2006}.
	
\subsection{Implications for LQC}
\label{LQCimplic}
Now we focus on the possible impact of the analysis above on LQC features. We have clarified how PQM implemented on the Ashtekar-Barbero-Immirzi connection implies a quantum cosmology which, similarly to the $\mu_0$ scheme of LQC, is characterized by a maximum energy density dependent on the quantum number $\overline{k}_{\phi}$. Differently, its implementation on the volume-like variable reproduces a bouncing dynamics with the same morphology of the $\bar{\mu}$ scheme of LQC. Despite this non-equivalence between the two formulations, in the previous subsection we have demonstrated that the semiclassical equations of motion can be mapped from the connection to volume-like variables by a redefinition of the discretization parameter as a function of the coordinates. Even if this issue of non-constant lattice step is not properly addressed in the full quantum polymer picture, this scenario can be analogous to the change of basis used to obtain the $\bar{\mu}$ scheme of LQC.
	
In the original formulation of LQC the dynamics of the isotropic Universe is described via the canonical couple $(p,c)$. The fundamental states of the theory, denoted by $\ket{\mu}$, are eigenstates of the momentum operator:
	\begin{equation}
		\hat{p}\,\ket{\mu}=\frac{1}{6}\,\ket{\mu}.
		\label{eg1}
	\end{equation}
While the operator $\hat{c}$ remains undefined, we can define a translational operator via the relation
	\begin{equation}
		\widehat{e^{\lambda c}}\,\ket{\mu}=\ket{\mu+\lambda}, 
		\label{eg2}
	\end{equation}
where $\lambda$ is a constant step. Now, the choice of the $\mu_0$ or $\bar{\mu}$ scheme amounts to giving the minimum area eigenvalue a kinematical or dynamical character \cite{Ashtekar_2006,BojowaldOriginalLQC,Ashtekar2_2006,Ashtekar_2011review,Ashtekar_2008}. In the $\bar{\mu}$ scheme of LQC the basic relation for the minimum area element is formulated as
	\begin{equation}
		\bar{\mu}^2\,p=\Delta, 
		\label{eg3}
	\end{equation}
where $\Delta$ is the Area Gap from full LQG, and it states the necessity to deal with physical values of the area spectrum properly scaled by the momentum (i.e. the squared scale factor).
	
Now, the translational operator would act on the states as
	\begin{equation}
		\widehat{e^{i\bar{\mu}(p)\,c}\,}\ket{\mu}=e^{\bar{\mu}(\mu)\,\dv{\mu}}\,\ket{\mu};
		\label{eg4}
	\end{equation}
its implementation becomes more natural when we change to the variable $v=p^\frac{3}{2}$, i.e. to a volume coordinate (this is immediate observing that $\bar{\mu}\propto\frac{1}{\sqrt{\mu\,}}$). This change of basis can be thought of as the passage from a translational operator with $p$-dependent step to a representation that defines a natural constant spacing in the $\mu$ space. This situation is similar to the case discussed in the polymer formulation, when in \eqref{beta'} the lattice step was promoted to a function of the generalized coordinate to restore the invariance of the semiclassical dynamics.
	
Since on a semiclassical level the polymer and LQC dynamics are comparable and we have shown that the physical properties of the Universe are dictated by the representation with a constant lattice step, we can infer an interesting feature of LQC. At least on a semiclassical level, the use of the volume coordinate is legitimated on the same level as the Ashtekar-Barbero-Immirzi connection (the privileged choice in LQG), because the universal value of the critical energy density (that is independent from $\overline{k}_{\phi}$), observed in the former case, would remain valid also for the evolution of the latter framework as long as we take into account the non-constant lattice parameter. In this respect, the proof of the complete equivalence of the $\mu_0$ and $\bar{\mu}$ schemes of LQC (clearly absent in the quantum polymer picture, as shown by our analysis, where a constant lattice step is considered) would require the non-trivial technical question of addressing the action of a translational operator with a coordinate-dependent step.
	
\section{Conclusions} \label{concl}
In this work we analyzed the dynamics of the isotropic Universe filled with a massless scalar field in the framework of PQC. We performed the semiclassical analysis for both the Ashtekar connection and the generalized coordinate conjugate to the volume, confirming the existence of a bouncing early Universe. The main conclusion is that the Big Bounce has different properties in the two sets of variables: an intrinsic cut-off emerges in the cosmological dynamics only in terms of the volume variable; the treatment in terms of the Ashtekar connection is still outlining a bouncing cosmology, but the scale of its manifestation depends on the initial conditions on the system. As a phenomenological comparison between the two schemes, we introduced in the semiclassical bouncing model a dissipative particle creation term in order to analyze the different behaviour of entropy over time and its dependence on the constant of motion $P_\phi$; we found that this effect is more relevant in the volume variables, while it is almost negligible in the Ashtekar ones. This discrepancy is more evident for small values of $P_\phi$, where the evolution in the two schemes is significantly different; indeed, in both cases the entropy production is strongly suppressed for large values of the initial conditions, corresponding to a scalar field energy density that becomes dominant over the dissipative radiation fluid.
	
We then proceeded to the full quantum analysis. We showed that, when using the Ashtekar connection, the expectation value of the energy density has the same behaviour of the semiclassical case and its maximum value is determined by the energy-like eigenvalue $\overline{k}_{\phi}$, while quantizing the volume variable yields an expectation value for the density operator dependent on fundamental constants only. As written above, a more thorough quantum description would require the study of variance and moments of the relevant operators \cite{BojowaldMoments}; furthermore, a possible future extension of this work could be to include inhomogeneities in the model and see whether PQC is affected by the same problems of LQC, such as the signature change due to the implementation of holonomy corrections \cite{SignChange,Bojowald_2015_1,Bojowald_2015_2} (even though in our framework these corrections are not present), but this goes beyond the scope of the present work.
	
In the discussion we showed how taking into account a polymer parameter depending on the momentum variable makes the equations of motion in the two settings equivalent, and this leads to a physically equivalent description of the cosmological Bounce in both the conjugate variables. Nonetheless, the nature of the Bounce is in any case fixed by the set of variables for which the polymer parameter is taken constant.
	
Our analysis in Sec. \ref{LQCimplic} suggested a semiclassical argument for the viability of the $\bar{\mu}$ scheme of LQC. In fact, if in such a formulation we interpret the change from the natural Ashtekar-Barbero-Immirzi connection to the volume variable as the restoration of a constant step in the translational operator from a momentum-depending one, we can apply the results obtained for the semiclassical equations of the polymer analysis, which mimic the semiclassical LQC dynamics. Therefore we can say that the universal critical energy density, coming from the choice of the volume base and its conjugate variable, is expected to remain constant also for the corresponding dynamics in terms of the natural $SU(2)$ connection, for which the translational operator has a non-constant step (we stress that the polymer configurational coordinate corresponds exactly to the variable $p$ of LQC, when the first order minisuperspace is considered). This consideration suggests that a further effort has to be made in this direction, from a quantum point of view, by solving the difficulties in implementing a translational operator associated to a  non-constant displacement.
	
\bibliography{bibl.bib}
	\begin{adjustwidth}{-\extralength}{0cm}
		\reftitle{References}
	\end{adjustwidth}
\end{document}